\documentclass{aa}  
\usepackage{graphicx}
\usepackage{txfonts}
\def\kms{km~s$^{-1}$}
\def\comain{$^{12}$CO~(1--0)}
\def\coisot{$^{13}$CO~(1--0)}

\newcommand{\um}{$\mu$m}

\newcommand{\cmdue}{cm$^{-2}$}

\newcommand{\msun}{M$_{\odot}$}
\newcommand{\firstcat}{$^{12}$CO catalogue}
\newcommand{\secondcat}{$^{13}$CO catalogue}

\def\herschel{{\it Herschel}}

\begin{document}

 \title{Molecular cloud catalogue from $^{13}$CO (1--0) data of the Forgotten Quadrant Survey\footnote{Table A.1 is only available in electronic form at the CDS  via anonymous ftp to cdsarc.u-strasbg.fr (130.79.128.5) or via http://cdsweb.u-strasbg.fr/cgi-bin/qcat?J/A+A/}}

\titlerunning{A molecular cloud catalogue from $^{13}$CO (1--0) data of the Forgotten Quadrant Survey}

   \author{M. Benedettini\inst{1}, A. Traficante\inst{1}, L. Olmi\inst{2}, S. Pezzuto\inst{1}, A. Baldeschi\inst{3,1}, S. Molinari\inst{1}, D. Elia\inst{1}, E. Schisano\inst{1}, M. Merello\inst{4}, F. Fontani\inst{2}, K. L. J. Rygl\inst{5}, J. Brand\inst{5}, M. T. Beltr\'{a}n\inst{2}, R. Cesaroni\inst{2}, S. J. Liu\inst{1}, L. Testi\inst{2,6}
          }

   \institute{
INAF -- Istituto di Astrofisica e Planetologia Spaziali, via Fosso del Cavaliere 100, 00133 Roma, Italy
\and
INAF -- Osservatorio Astrofisico di Arcetri, Largo E. Fermi 5, 50125, Firenze, Italy 
\and
Center for Interdisciplinary Exploration and Research in Astrophysics and Department of Physics and Astronomy, Northwestern University, Evanston, IL 60208
\and
Departamento de Astronom\'ia, Universidad de Chile, Casilla 36-D, Santiago, Chile
\and
INAF -- Istituto di Radioastronomia \& Italian ALMA Regional Centre, via P. Gobetti 101, 40129, Bologna
\and
ESO/European Southern Observatory, Karl-Schwarzschild-Strasse 2, 85748, Garching bei M\"{u}nchen, Germany
             }
\authorrunning{Benedettini et al.}

   \date{Received ; accepted}

 
  \abstract
   {New-generation spectroscopic surveys of the Milky Way plane have been revealing the structure of the interstellar medium, allowing the simultaneous study of dense structures from single star-forming objects or systems to entire spiral arms.}
   {The good sensitivity of the new surveys and the development of dedicated algorithms now enable building extensive catalogues of molecular clouds and deriving good estimates of their physical properties. This allows studying the behaviour of these properties across the Galaxy.}
   {We present the catalogue of molecular clouds extracted from the $^{13}$CO (1--0) data cubes of the Forgotten Quadrant Survey, which mapped the Galactic plane in the range 220\degr$<l<$240\degr\, and -2\fdg5$<b<$0\degr\  in $^{12}$CO~(1--0) and $^{13}$CO~(1--0). We compared the properties of the clouds of our catalogue with those of other catalogues.}
   {The catalogue contains 87 molecular clouds for which the main physical parameters such as area, mass, distance, velocity dispersion, and virial parameter were derived. These structures are overall less extended and less massive than the molecular clouds identified in the $^{12}$CO~(1--0) data-set because they trace the brightest and densest part of the $^{12}$CO~(1--0) clouds. Conversely, the distribution of aspect ratio, equivalent spherical radius, velocity dispersion, and virial parameter in the two catalogues are similar. 
   The mean value of the mass surface density of molecular clouds is 87$\pm$55 M$_{\odot}$ pc$^{-2}$ and is almost constant across the galactocentric radius, indicating that this parameter, which is a proxy of star formation, is mostly affected by local conditions.}
   {In data of the Forgotten Quadrant Survey, we find a good agreement between the total mass and velocity dispersion of the clouds derived from $^{12}$CO~(1--0) and $^{13}$CO~(1--0). This is likely because in the surveyed portion of the Galactic plane, the H$_2$ column density is not particularly high, leading to a CO emission with a not very high optical depth. This mitigates the effects of the different line opacities between the two tracers on the derived physical parameters. This is a common feature in the outer Galaxy, but our result cannot be readily generalised to the entire Milky Way because regions with higher particle density could show a different behaviour.}
   

   \maketitle
%
\section{Introduction}

In the past decades, the advent of new spectrometers mounted at the focal plane of large radio antennas has allowed us to efficiently carry out new spectral surveys of the molecular gas in the Galactic plane in CO and its isotopologues. Some of these large programs are 
the Galactic Ring Survey (GRS, \citealt{jackson2006}), the Exeter-FCRAO CO  Galactic Plane Survey \citep{mottram2010}, the Mopra Southern Galactic Plane CO survey \citep{burton2013}, The Milky Way Image Scroll Painting (MWISP, \citealt{jiang2013}), the CO High-Resolution Survey (COHRS, \citealt{dempsey2013}), the Three-mm Ultimate Mopra Milky Way Survey \citep{barnes2015}, the CO Heterodyne Inner Milky Way Plane Survey (CHIMPS, \citealt{rigby2016}), the Structure, Excitation, and Dynamics of the Inner Galactic Inter-Stellar Medium  survey (SEDIGISM, \citealt{schuller2017}), and the FOREST Unbiased Galactic plane Imaging survey with the Nobeyama 45m telescope (FUGIN, \citealt{umemoto2017}). We contributed to this general effort with the Forgotten Quadrant Survey (FQS, Benedettini et al. 2020, hereafter \citealt{benedettini2020}). The FQS is an ESO project that used the Arizona Radio Observatory (ARO) 12m antenna to map a portion of the midplane of the third quadrant of the Milky Way, in the range 220\degr $< l < $ 240\degr\, and -2\fdg5 $ < b  < $ 0\degr\  in \comain\, and \coisot at a spectral resolution of 0.65 \kms\, and 0.26 \kms, respectively. All these CO surveys with their good sensitivity have allowed us to detect the molecular component of the diffuse interstellar medium (ISM) and to characterise its distribution in the Galactic plane because an estimate of the kinematical distance could be derived from their data. At the same time, dedicated algorithms, for instance, CLUMPFIND developed by \cite{williams1994}, CPROPS \citep{rosolowsky2006}, or SCIMES \citep{colombo2015}, have increased the capability of correctly identifying coherent structures in these high spectral resolution data and of deriving good estimates of their physical properties with which extended catalogues of molecular clouds (MCs) were built. This large database now allows reconstructing the distribution of molecular gas in our Galaxy and studying how the diffuse ISM gathers to form increasingly denser structures, from the spiral arms traced by the giant molecular clouds to clumps and cores, where new stars with their planetary systems form.

In \citet{benedettini2020}, we presented a catalogue of MCs extracted from the \comain\, spectral cubes (hereafter the \firstcat) that described how the molecular gas is organised in the surveyed portion of the outer Galaxy. With this new-generation data of improved quality, we were able to detect many more clouds than in previous data \citep{dame2001}. In particular, we were able to identify not only the typical giant molecular clouds with sizes of tens of parsec, but also the small clouds and we were able to resolve the internal cloud structure at the sub-parsec scale up to a distance of a few kiloparsec. The FQS \firstcat\, contains 263 MCs grouped in three main structures that correspond to the Local, Perseus, and Outer arms, up to a distance of $\sim$8.6 kpc from the Sun. Part of the area surveyed by FQS, corresponding to the CMa OB1 complex, was also observed by the MWISP survey by \citet{lin2021}, who compared the three largest clouds at a distance of about 1 kpc. They revealed a difference of physical properties, evolutionary stages, and levels of star formation activity in the three subregions of the CMa OB1 complex.

In this paper we add another tile to the mosaic by presenting the MC catalogue extracted from the \coisot\, data cubes of the FQS survey. The structure of the paper is as follows. Section 2 describes the observational setup and the structure of the final products. Section 3 presents the catalogue of MCs extracted from the FQS \coisot\, spectral data cubes. In Sect. 4 we describe the three different methods we used to estimate the mass of the clouds and compare the results. The physical parameters of the MCs are analysed and compared  with those of the FQS \firstcat\, in Sect. 5. In Sect. 6 we analyse the behaviour of the cloud mass surface density as function of the galactocentric radius. The summary and main conclusions are presented in Sect. 7. 

\section{Observations and data reduction}

We used the ARO AEM ALMA prototype 12m antenna with the ALMA type band-3 receiver to map the Milky Way plane in the range 220\degr $ < l < $ 240\degr\, and -2\fdg5 $ < b < $ 0\degr, following the Galactic warp, in \comain\, and \coisot. The observational setup and the data reduction pipeline were described in \citet{benedettini2020}. Here we briefly recall that the receiver was tuned to 115.271 GHz and 110.201 GHz for \comain\, and \coisot, respectively. The backend was a 256-channel filter bank at 250 kHz spectral resolution (hereafter FB250), corresponding to a total velocity coverage of 166 \kms\, and a velocity resolution of 0.65 \kms, in parallel with a second 256-channel filter bank at 100 kHz spectral resolution (hereafter FB100), corresponding to a total velocity coverage of 66 \kms\, and a velocity resolution of 0.26 \kms. Raw data are reduced with a dedicated pipeline described in \citet{benedettini2020}. The final products are spectral data cubes for  \comain\, and \coisot, with channels of 0.3~\kms\  from the FB100 backend, and with 1~\kms\, channels from the FB250 backend. The spatial pixel size for both cubes is 17\farcs3, that is, about one-third of the telescope beam. The median root mean square (rms) noise of the main-beam temperature at pixel level for the FB250 cubes is 0.53 K and 0.22 K for \comain\, and \coisot, respectively. 

\begin{figure*}
   \centering
    \includegraphics[width=19.cm]{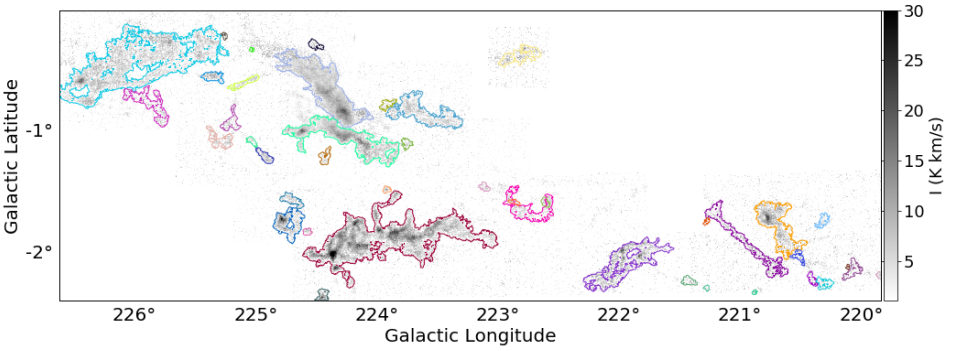}
    \includegraphics[width=19.cm]{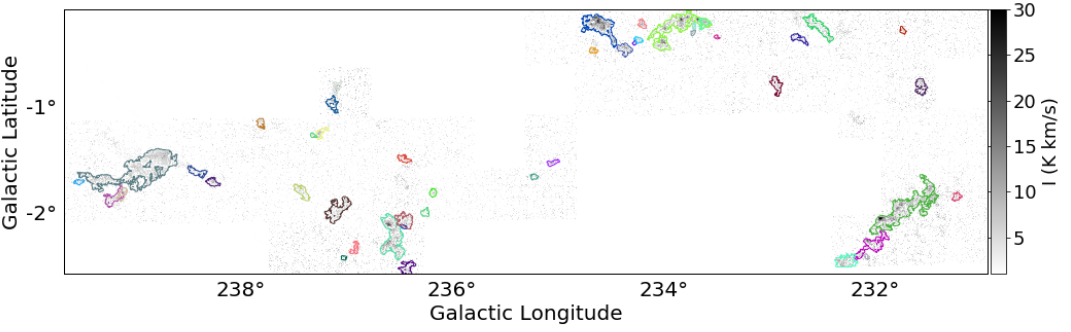}
    \caption{Edge of the MCs of the \secondcat\  drawn over the integrated intensity map of \coisot\, (intensity bar on the right).}
   \label{fig:clouds}
\end{figure*}

\section{Molecular cloud catalogue from \coisot\, data}

We used the \coisot\, data cubes to produce a catalogue of MCs, in analogy to the \comain\, line MCs catalogue presented in \citet{benedettini2020}. In our FQS data, the emission of the \coisot\,line is less extended than that of the \comain\, line. As a consequence, we do not detect the low-column density layers of the \comain\, clouds in the \coisot\, data, and structures that appear connected in \comain\, can instead be composed of smaller separate clouds in the \coisot\, catalogue. On the other hand, the \coisot\, line has a lower optical depth, typically $<$1 in our data (see Sect. \ref{sect:cloud_mass}). It therefore is a better tracer of the total CO column density than the optically thick $^{12}$CO. Moreover, several MC catalogues in the literature were derived from \coisot\, data, and this new catalogue will allow us to compare our results with a more homogeneous data-set. 

To identify the clouds, as in \citet{benedettini2020}, we used the algorithm called Spectral Clustering for Interstellar Molecular Emission Segmentation ({\it SCIMES}) \citep{colombo2015}, which is based on a cluster analysis of dendrograms of 3D ({\it l,b,$\varv$}) data cubes. We applied {\it SCIMES} to the \coisot\, spectral data cubes from the FB250 filter bank. In order to select well-defined structures and reduce contamination from noisy regions, we masked the cubes by selecting only pixels in which the signal is higher than 5$\times$rms in two consecutive velocity channels. We further extended the mask to include all adjacent pixels in which the signal is higher than 2$\times$rms. We then produced the dendrogram tree of the masked cube.  We set the minimum signal in the pixel (min\_value) to 0.61 K (equivalent to 3$\times$rms), the minimum difference between two peaks for being considered separate leaves (min\_delta) to 8$\times$rms, and the minimum number of pixels of the spectral cube needed for a leaf to be independent (min\_npix) to those contained in three times the telescope-beam solid angle.  With respect to the set of the {\it SCIMES} parameters used in \citet{benedettini2020}, we increased the min\_value from 2$\times$rms to 3$\times$rms to select only pixels with a well-defined line emission that guarantees a good line fitting and min\_delta from 7$\times$rms to 8$\times$rms to reduce the level of fragmentation of the detected structures. We finally applied the spectral clustering using the volume of the isosurface in the ({\it l,b,$\varv$}) space as a criterion. Because we were also interested in clouds that may simply have little substructure within them, we also included the single leaves of the dendrogram that had been excluded by the clustering algorithm. 

In total, we identified 87 MCs in our \coisot\, FB250 data cubes. Sixty-seven MCs are composed of a single leaf, indicating that they are coherent structures. A view of the position and contours of the identified structures is shown in Fig. \ref{fig:clouds}. For each cloud, we measured the following properties, using the definition in \citet{rosolowsky2006}: centroid position in Galactic coordinates; semi-major ($\sigma_{\rm maj}$) and semi-minor ($\sigma_{\rm min}$) axes, estimated as the rms of the intensity-weighted second moments along the direction of maximum cloud extent and its perpendicular direction, respectively; position angle; intensity-weighted velocity; and relative velocity dispersion ($\sigma_{v}$). From the geometrical mean of the semi-axes, we derived the radius of the equivalent spherical cloud as $R = 1.91\sqrt{\sigma_{\rm maj}\sigma_{\rm min}}$ \citep{rosolowsky2006}. 

From the central velocity of the \coisot\, line, we derived the kinematic distance of the emitting cloud by applying a Galaxy rotation model. We used the IDL routine of the CPROPS package \citep{rosolowsky2006} for the distance estimate, assuming a flat Galactic rotation curve, which is a good approximation for the outer Galaxy, with the solar galactocentric radius $R_0$ = 8.34 kpc and rotation velocity $\Theta_0$ = 240 \kms\, \citep{reid2014}. Because we observe in  directions pointing outside of the solar circle, our distance estimate is not affected by the near and far distance ambiguity. We also converted the heliocentric distance into galactocentric radius $R_{\rm gal}$. The mass of the clouds was estimated with three different methods that are described in Sect. \ref{sect:cloud_mass}. To be consistent with \cite{benedettini2020}, we used the mass derived from the \herschel\, H$_2$ column density map (Sect. \ref{sect:mass_dust}) to calculate the average mass surface density ($\Sigma$) of the clouds and the virial parameter, $\alpha_{\rm vir} = \frac{5 \sigma_{\rm v}^2 R}{M G}$ \citep{mckee1992}, where $G$ is the gravitational constant. We derived the velocity dispersion of the gas from the measured full width at half maximum (FWHM) of the \coisot\, line through the relation $\sigma_{\rm v}  = FWHM / \sqrt{8 \ln(2)}$. We highlight that this is a good estimate of the velocity dispersion because the lines are spectrally resolved, allowing a good measure of the FWHM, and its optical depth is low, with values at the line peak below 3 in 98\% of the spatial pixels of the MCS. Moreover, we observe in the direction of a portion of the outer Galaxy in which the cloud crowding is not an issue. In this region the \coisot\, observed lines as well as \comain\, \citep{benedettini2020} are well fitted with a single-Gaussian profile, and our measurement of the gas velocity dispersion of the single MCs is not affected by blending of emission of possible multiple clouds along the same line of sight. 

All the derived parameters are reported in Table \ref{tab:MC} for a sub-sample of MCs. The complete table is available in electronic form at the CDS \footnote{CDS is available via anonymous ftp to cdsarc.u-strasbg.fr (130.79.128.5) or via http://cdsweb.u-strasbg.fr/cgi-bin/qcat?J/A+A/}. In the catalogue, we also give the number of leaves identified by the dendrogram that compose the cloud. Because the \coisot\, map is larger than the \herschel\, H$_2$ column density map, which does not cover Galactic latitudes below $\sim$-2\degr, some clouds identified in the \coisot\, map are only partially covered or not covered at all by the \herschel\, H$_2$ column density map. The mass and surface density derived for these clouds are lower limits and the virial parameter is an upper limit; these clouds are flagged with flag1 = 1 in Table \ref{tab:MC}. Other clouds extend up to the edge of the \coisot\, map and might therefore not be completely mapped by our observations. For these clouds, identified with flag2 = 2 in Table \ref{tab:MC}, the measured parameters have an uncertainty that depends on how much CO emission was missed.

\section{Mass estimate of molecular clouds}
\label{sect:cloud_mass}

The mass of MCs is derived by integrating the H$_2$ column density over the projected area in the sky with the relation
\begin{equation}
\label{eq:mass}
 M = \mu_{\rm H_2} m_{\rm H} \int{N{\rm (H_2)} {\rm d}A}
,\end{equation}
where $\mu_{\rm H_2}$ = 2.8 is the mean molecular weight for a hydrogen molecule that takes into account the presence of helium, $m_{\rm H}$ is the mass of the hydrogen atom, $A$ is the area of the cloud, and $N{\rm (H_2)}$ is the molecular hydrogen column density. 
For the clouds of the FQS catalogue, we have different possibilities to derive $N{\rm (H_2)}$ and, consequently, the mass, either by using the cold dust as a tracer, exploiting the \herschel\, data, as we did in \citet{benedettini2020} for the MC catalogue derived from \comain\, emission, or by using the molecular gas as a tracer, exploiting the FQS data. In the following subsections, we describe three different methods that we used to estimate the mass of the clouds, and we compare the results.

\subsection{Mass from the $^{13}$CO column density}
\label{sect:cd_co}

The FQS data-set offers the possibility of deriving the mass of the clouds from the \coisot\, line emitted from the molecular gas by deriving the $^{13}$CO column density and converting it into H$_2$ column density using the $^{13}$CO chemical abundance with respect to H$_2$.  We call this mass $M_{N(^{13}\rm CO)}$. We used the mask of the \coisot\, FB250 spectral cube produced by the {\it SCIMES} algorithm for the identification of the MCs to define the spatial and spectral range pertinent to each MC, and we derived the H$_2$ column density map for each MC of the catalogue. We assumed that the emission fills the beam, which is good for the extended emission of gas in MCs, and a uniform excitation temperature within the beam, equal for both \comain\, and \coisot. This latest assumption is formally correct in local thermal equilibrium (LTE) conditions, although density gradients and non-LTE conditions in local portion of the clouds can lead to differences in excitation temperature of \comain\, and \coisot. The excitation temperature $T_{\rm ex}$ was derived from the \comain\, brightness temperature in each ({\it l,b,$\varv$}) pixel under the assumption that the \comain\, emission is optically thick, which is usually the case for MCs, with the formula

\begin{equation}
 T_{\rm ex} = \frac{5.53}{{\rm ln}[1 + 5.53/(T(^{12}{\rm CO}) + 0.837)]} ~[K]
,\end{equation}
where $T(^{12}{\rm CO})$ is the main-beam brightness temperature of \comain\, at the ({\it l,b,$\varv$}) pixel. This equation includes the subtraction of the cosmic microwave background at $T=$ 2.73 K. The excitation temperature at the peak of the line in the mapped region ranges between 3.6 K and 30.8 K, with a median value of 9.5 K.

The optical depth of the \coisot\, was calculated in each ({\it l,b,$\varv$}) pixel from the formula
\begin{equation}
 \tau_{13} = -{\rm ln}\left[ 1 - \frac{T(^{13}{\rm CO})]/5.3}{(e^{5.3/T_{\rm ex}} - 1)^{-1} - 0.16}\right]
 ,\end{equation}
where $T(^{13}{\rm CO})$ is the main-beam brightness temperature of \coisot\, in the pixel. We found that the median value of $\tau_{13}$ for all the pixels is 0.2 (0.6 at the peak of the line), with the 5th$^{}$ percentiles of 0.03 and the 95th$^{}$ percentiles of 0.9, indicating that in this portion of the Galactic plane, at the spatial scales probed by our observations (larger than about 0.27 pc for distances larger than 1 kpc), the \coisot\, line is almost always optically thin or moderately thick. It is therefore able to trace the largest part of the molecular gas in the clouds, from which a meaningful estimate of the cloud mass can be derived.

We derived the \coisot\, column densities at each ({\it l,b}) pixel with the following formula:
\begin{equation}
 N(^{13}{\rm CO}){\rm [cm^{-2}]} = 2.6 \times 10^{14} \int{\frac{\tau_{13}T_{\rm ex}}{1-e^{\frac{-5.3}{T_{\rm ex}}}}  {\rm d}\varv ~ [\rm K ~km ~s^{-1}]}
.\end{equation} 
The \coisot\, column density map was then converted into H$_2$ column density map by assuming a relative chemical abundance of [$^{13}$CO]/[H$_2$] = 1.4$\times$10$^{-6}$ \citep{frerking1982}. The mass of the MCs was calculated with Eq. \ref{eq:mass}.

\subsection{Mass from the integrated \comain\, line emission}
\label{sect:mass_xco}

A method that is commonly used in the literature to derive the H$_2$ column density when only the \comain\, line is available is to use the empirical relation between $N$(H$_2$) and the integrated intensity of the \comain\, line, $I(^{12}{\rm CO})$, that is,
\begin{equation}
\label{eq:xco}
 N({\rm H_2}) = X(^{12}{\rm CO}) \,  I(^{12}{\rm CO}),\end{equation}
where $X(^{12}{\rm CO})$ is the CO-to-H$_2$ conversion factor. Many studies have been conducted to derive this factor with different methods. In this paper, we assumed the widely used value $X(^{12}{\rm CO}) = 2\times 10^{20}$ cm$^{-2}$ (K km s$^{-1}$)$^{-1}$, which has been recommended by \citet{bolatto2013} in his review paper.  We call the mass derived with this method $M_{X(^{12}{\rm CO})}$. In order to correctly compare this mass with the one derived in the previous section, we integrated the \comain\, line emission in the same area of the cloud, that is, the one defined by the \coisot\, emission. Recent papers have shown that when measured in new-generation CO surveys, with arcminute or sub-arcminute spatial resolution, the CO-to-H$_2$ conversion factor spans more than one order of magnitude from pixel to pixel (e.g. \citealt{barnes2015,schuller2017,benedettini2020}) due to local differentiation of the physical conditions between the clouds and within the same MC when it is spatially well resolved. However, in general, the mean value agrees within the error with the standard value.

\subsection{Mass from the FIR dust emission}
\label{sect:mass_dust}
The third option to derive the mass of the MCs of the FQS catalogue, which we applied in \citet{benedettini2020}, is to use the H$_2$ column density derived from the \herschel\, photometric data. We refer to this mass as $M_{\rm dust}$. The \herschel\, H$_2$ column density maps were derived from the fitting of the spectral energy distribution (SED) of the cold dust in the wavelength range from 160 \um\, to 500 \um\,  \citep{schisano2020},  under the assumption of a gas-to-dust ratio of 100 and a dust opacity law of $\kappa_{\lambda}=\kappa_{\rm 300} ~ (\lambda/300~\mu m)^{-\beta}$ with $\kappa_{\rm 300}$=0.1 cm$^2$g$^{-1}$ and a grain emissivity parameter $\beta$=2 \citep{hildebrand1983}. The \herschel\, maps have a spatial resolution of 36\arcsec, which is similar to that of the FQS data, and they were convolved and regridded to match the FQS products.
Being derived from photometric data, \herschel\, H$_2$ column density map cannot disentangle clouds at different distances that overlap along the same line of sight. In our FQS \secondcat, only eight MCs have a small spatial overlap with another cloud, as shown in Fig. \ref{fig:clouds}. In the overlapping spatial pixels, we distributed the total H$_2$ column density of the \herschel\, map among the two cospatial clouds in proportion to their \coisot\, intensity. This is a reasonable approach because the \coisot\, emission is optically thin in our data-set, as shown in Sect. \ref{sect:cd_co}, and in this case, the emission is proportional to the column density.

\begin{figure*}
   \centering
    \includegraphics[width=\textwidth]{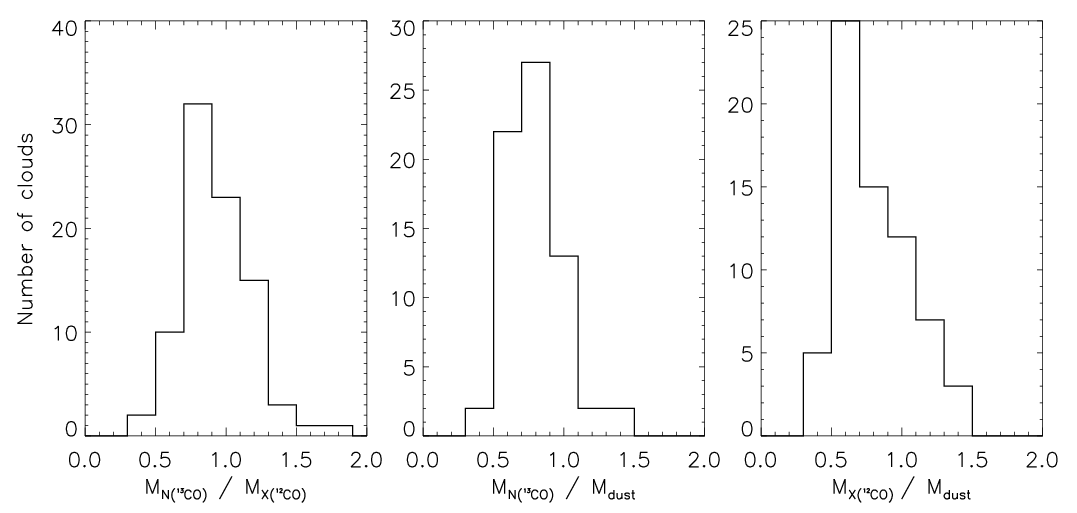}
    \caption{Histograms of the ratios of the cloud mass estimates with three different methods. {\it Left:} Ratio of the mass derived from the $N(^{13}\rm CO)$ column density and that from $X(^{12}{\rm CO})$. {\it Centre:} Ratio of the mass derived from the $N(^{13}\rm CO)$ and that from the \herschel-based H$_2$ column density, which traces the cold dust. {\it Right:} Ratio of the mass derived from $X(^{12}{\rm CO})$ and that from the \herschel-based H$_2$ column density.}
   \label{fig:mass}
\end{figure*}

\begin{table*}
\caption{Ratio of the masses of the MCs of the \firstcat\, derived with different methods.}             
\label{tab:mean}      
\centering                          
\begin{tabular}{l c c c }        
\hline\hline                 
 mass ratio   & median & 10$^{th}$ percentile & 90$^{th}$ percentile \\ 
\hline                        
$M_{N(^{13}\rm CO)}$ / $M_{X(^{12}{\rm CO})}$  & 0.99 & 0.76 & 1.36 \\
$M_{N(^{13}\rm CO)}$ / $M_{\rm dust}$   & 0.86 & 0.66 & 1.15 \\
$M_{X(^{12}{\rm CO})}$ / $M_{\rm dust}$   & 0.86 & 0.61 & 1.34 \\
\hline                                   
\end{tabular}
\end{table*}

\subsection{Comparison of the three mass estimates}

The masses of the MCs of the FQS catalog derived with the three methods are reported in Table \ref{tab:MC}. We point out that the three masses are calculated in the same projected area onto the plane of the sky for each cloud. In Fig. \ref{fig:mass}, we show the distribution of their ratios, $M_{N(^{13}\rm CO)}$ / $M_{X(^{12}{\rm CO})}$, $M_{N(^{13}\rm CO)}$ / $M_{\rm dust}$ , and $M_{X(^{12}{\rm CO})}$ / $M_{\rm dust}$, and in Table \ref{tab:mean}, we report the median values and the 10th$^{}$ and 90th$^{}$ percentiles of these ratios. In general, we find a good agreement between the mass estimates, with ratios ranging from 0.4 to 1.9. In particular, the masses derived from CO data with $N(^{13}\rm CO)$ and $X(^{12}{\rm CO})$ are very similar, with a narrow distribution of their ratio centred at the median value of 0.99. This indicates that at the sensitivity level of our survey, the two lines \comain\, and \coisot \ in general trace a similar column of gas. \coisot\,  loses only the more tenuous outskirts of the clouds, which contribute little to the total cloud mass. Moreover, the similarity of these two mass estimates indicates that despite the high variability of the CO-to-H$_2$ conversion factor at the pixel level \citep{benedettini2020}, the assumption of a constant value for $X(^{12}{\rm CO})$ is a valid method for deriving the total mass in extended structures as the MCs, at least in this portion of the Milky Way. On the other hand, for the majority of the MCs, the mass derived from the CO gas is lower than that derived from the cold dust, with a median value of 0.86 for both  $M_{N(^{13}\rm CO)} / M_{\rm dust}$ and $M_{X(^{12}{\rm CO})} / M_{\rm dust}$. The same trend was also found by other authors who analysed CO data collected with different telescopes (NANTEN, \citealt{elia2013} and MOPRA, \citealt{olmi2016}) in some of the regions of the Galactic plane that are mapped in FQS. This is likely because the cold dust at the far-infrared wavelengths of the \herschel\, observations is a truly optically thin tracer that is more effective in tracing all the material along the line of sight than CO because \comain\, is usually optically thick in MCs and also \coisot, in spite of its lower opacity, becomes optically thick in the densest regions such as clumps and cores. 

\begin{figure*}
   \centering
    \includegraphics[width=19.cm]{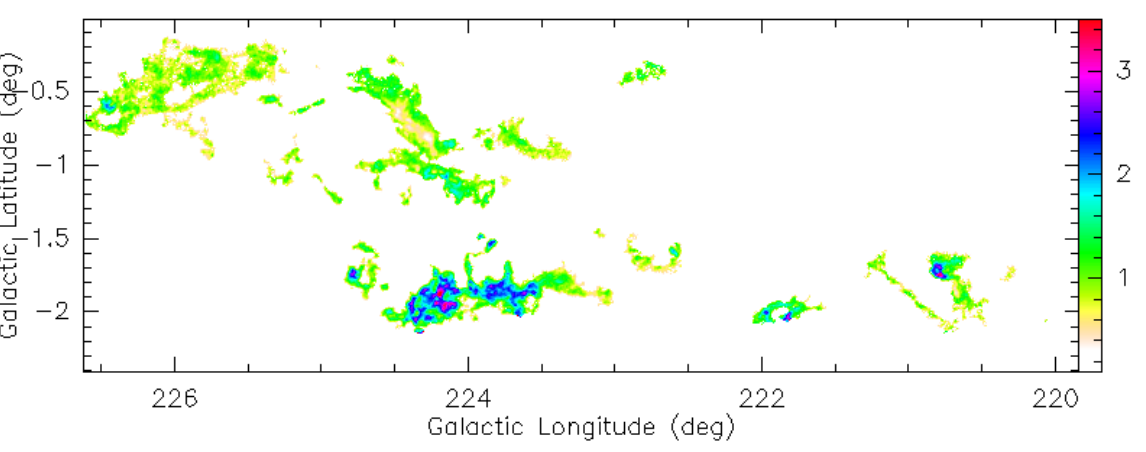}
    \includegraphics[width=19.cm]{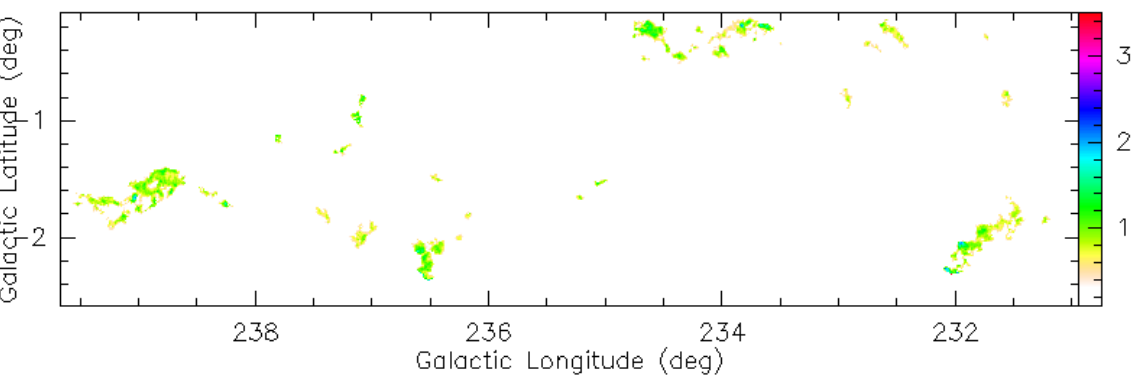}
    \caption{$N{(\rm H_2})_{N(^{13}\rm CO)} / N{(\rm H_2})_{\rm dust}$ ratio (right colour bar) in the $l$ = 220\degr--226\fdg6 field (top) and that in the $l$ = 231\degr--239\fdg7 field (bottom).}
   \label{fig:nh2pixels}
\end{figure*}

\begin{figure*}
   \centering
    \includegraphics[width=0.9\textwidth]{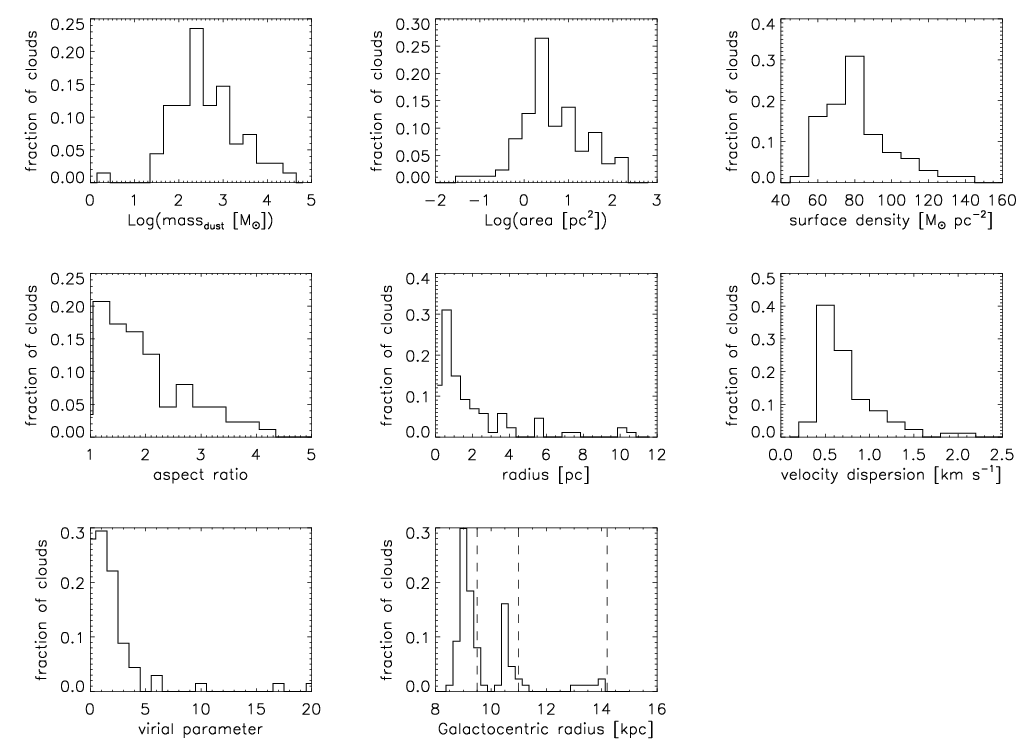}
   \caption{Histograms of the distributions of the derived physical properties for the MCs of the FQS \secondcat. {\it Top left:}  Masses derived from cold dust emission. {\it Top centre:} Area. {\it Top right:} Average mass surface density. {\it Middle left:} Aspect ratio ($\sigma_{\rm maj}$ / $\sigma_{\rm min}$). {\it Middle centre:} Equivalent spherical radius. {\it Middle right:} Velocity dispersion. {\it Bottom left:} Virial parameter. {\it Bottom centre:} galactocentric radius. The vertical dashed lines indicate the reference position of the Local Arm \citep{xu2013}, Perseus Arm \citep{choi2014}, and Outer Arm \citep{hou2014} at the central Galactic longitude of the surveyed region $l$ = 230\degr.}
   \label{fig:histograms}
   \end{figure*}

When comparing mass estimates, we need to take the sources of uncertainty into account that come from the assumptions that we made in deriving masses. On the one hand, for the CO-based mass estimates, we assumed a constant CO abundance for all the clouds, while region-to-region variations can occur depending on the chemical evolutionary stage and on the local physical conditions. We also assumed the same excitation temperature for  \comain\, and \coisot, while density gradients and non-LTE conditions can lead to differences in the excitation temperature of the two lines. On the other hand, for the dust-based mass estimates, we assumed a homogeneous dust opacity law, but differences within regions can exist due to the different properties of the dust grains and their state of growth. Moreover, there is evidence that the dust opacity depends on the H$_2$ column density (e.g. \citealt{martin2012,roy2013}) and that the grain emissivity parameter $\beta$, which gives the dependence of the opacity on the frequency, is anti-correlated with the temperature (e.g. \citealt{dupac2003,desert2008,juvela2015}). \citet{schisano2020} have shown that assuming a value of $\beta$ = 1.8 (which should be more appropriate for regions with higher temperature between $\sim$ 15 -- 20 K) instead of $\beta$ = 2 (which should be more appropriate for colder regions with temperature between $\sim$ 10 --15 K) has the effect of decreasing the column density by about 20\% and increasing the  temperature by about 6\%. Additionally, \citet{pezzuto2021} have shown that when the temperature of the dust is higher than 20 K, the SED fitting of the four Herschel bands with $\lambda \geq$ 160 \um underestimates the true dust temperature, and as a consequence, overestimates the H$_2$ column density. 
Finally, there are some indications that the gas-to-dust ratio decreases with galactocentric radius \citep{giannetti2017}. However, in our set of data, we do not find any correlation of the $M_{N(^{13}\rm CO)}$ / $M_{\rm dust}$ ratio with the galactocentric radius. This suggests that either there is no significant variation of the CO chemical abundance and the gas-to-dust ratio with distance in the first $\sim$ 6 kpc of the surveyed slice of the outer Galaxy or, less likely, that both the CO chemical abundance and the gas-to-dust ratio change with distance with the same trend. However, we must point out that the statistics at the larger distance is poor because our sample only includes six MCs with $R_{\rm gal} >$ 11.3 kpc (see the last panel of Fig. \ref{fig:histograms}). We do not find any correlation either of the ratio of the mass derived from CO and from dust with the integrated intensity of \comain\, and with the dust temperature.

In order to further explore the possible presence of the effects listed above in specific regions, we compared pixel-by-pixel the $N{(\rm H_2})$ maps derived from \coisot\, FQS data ($N{(\rm H_2})_{N(^{13}\rm CO)}$) with those derived from the SED fitting of the \herschel\, data ($N{(\rm H_2})_{\rm dust}$). The $N{(\rm H_2})_{N(^{13}\rm CO)}$ map was built by stacking all the maps of the 87 MCs of the catalogue.
In Fig. \ref{fig:nh2pixels} we show the pixel-by-pixel $N{(\rm H_2})_{N(^{13}\rm CO)} / N{(\rm H_2})_{\rm dust}$ ratio. We find that 65$\%$ of the pixels have $N{(\rm H_2})_{N(^{13}\rm CO)} / N{(\rm H_2})_{\rm dust}<$ 1, while only a small portion of the mapped area (4$\%$ of pixels) has $N{(\rm H_2})_{N(^{13}\rm CO)} / N{(\rm H_2})_{\rm dust}>$2. 
We note that the pixels with this high ratio belong to only four MCs: FQS-MC220.716-1.783 (index = 8) at ($l\simeq$220.7\degr, $b\sim$-1.8\degr), FQS-MC221.959-2.092 (index = 14) at ($l\simeq$221.9\degr, $b\sim$-2.1\degr), FQS-MC223.966-1.904 (index = 24) at ($l\simeq$224\degr\,$b\sim$-1.9\degr), and FQS-MC224.750-1.769 (index = 32) at ($l\simeq$224.8\degr, $b\sim$-1.8\degr).  These are the regions with the brightest CO emission (Figs. 6 and 8 in \citet{benedettini2020}) and the highest CO excitation temperature ($T_{\rm ex}\geq$ 12 K) as well as high dust temperature, suggesting that the presence of warm gas could affect the observed ratio. In MCs with index 8, 14, and 32, only a small portion of their projected area has $N{(\rm H_2})_{N(^{13}\rm CO)} / N{(\rm H_2})_{\rm dust}>$2, while in the cloud FQS-MC223.966-1.904,  the H$_2$ column density derived from CO in most of the area is higher than that derived from the dust. This is the MC with the brightest \comain\, emission in the surveyed area, and it also appears bright in the \herschel\, maps. This MC is part of the CaM OB1 association and roughly corresponds to the Sh 2-296 nebula, which is rich in gas ionised by the close-by O-type stars \citep{fernandes2019}. Two \ion{H}{II} regions are present in this area, S296 and S292 \citep{sharpless1959}. For this region, both the $T_{\rm ex}$ estimated from \comain\, and the dust temperature derived from the SED fitting of the \herschel\, fluxes are the highest of the whole surveyed region, with values between $\sim$ 15 K and $\sim$ 29 K, respectively. At this temperature, the mass derived from \herschel\, data is likely underestimated, which makes the measured $N{(\rm H_2})_{N(^{13}\rm CO)} / N{(\rm H_2})_{\rm dust}$ an upper limit. This high temperature and the presence of ionised gas may modifiy the chemical abundance and/or different properties of the dust grains in this particular cloud, which might explain the high measured $N{(\rm H_2})_{N(^{13}\rm CO)} / N{(\rm H_2})_{\rm dust}$ ratio. A substantial modification of the chemical abundance ratio of $^{13}$CO and C$^{18}$O in the region that corresponds to our MC FQS-MC223.966-1.904 with respect to its adjacent clouds was found by \citet{lin2021} as well.

\begin{table}
\caption{Statistical physical properties of the MCs of the FQS catalogue derived from the \coisot\, data.}             
\label{tab:statistic}      
\centering                          
\begin{tabular}{l c c}        
\hline\hline                 
    & median & mode \\    
\hline                        
mass$_ {\rm dust}$ (\msun)     & 443 & 251 \\
area (pc$^2$)    & 4.7 & 2.5 \\
surface density (\msun\, pc$^{-2}$) & 86 & 80 \\
aspect ratio     & 2.0 & 1.2 \\
equivalent spherical radius (pc) & 1.4 & 0.6 \\
velocity dispersion (\kms) & 0.7 & 0.5 \\
virial parameter & 1.87 & 1.00 \\
heliocentric distance (kpc)   & 1.57 & 0.89 \\
galactocentric radius (kpc)   & 9.33 & 9.00 \\
\hline                                   
\end{tabular}
\end{table}   

\section{Analysis of the \coisot\, molecular cloud catalogue}

\subsection{Global properties}

In Fig. \ref{fig:histograms} we show the histograms of some of the physical parameters of the MCs, and in Table \ref{tab:statistic} we list the median and modal values of these parameters. The MCs are distributed along a heliocentric distance from 0.39 kpc to 7.74 kpc, which corresponds to a galactocentric radius from 8.64 kpc to 14.15 kpc. The clouds are gathered into three groups that correspond to the three spiral arms of the Milky Way (Local, Perseus, and Outer arms), which are well visible in the last panel of Fig. \ref{fig:histograms}. Because of the different distances, the minimum detectable size of the clouds changes significantly in the three arms. In particular, all the MCs of the catalogue have a projected area on the sky that is larger than twice the area of the telescope beam, which corresponds to a minimum area from 0.16 pc$^2$ at 1 kpc to 10.31 pc$^2$ at 8 kpc.

\begin{figure}
   \centering
    \includegraphics[width=9.cm]{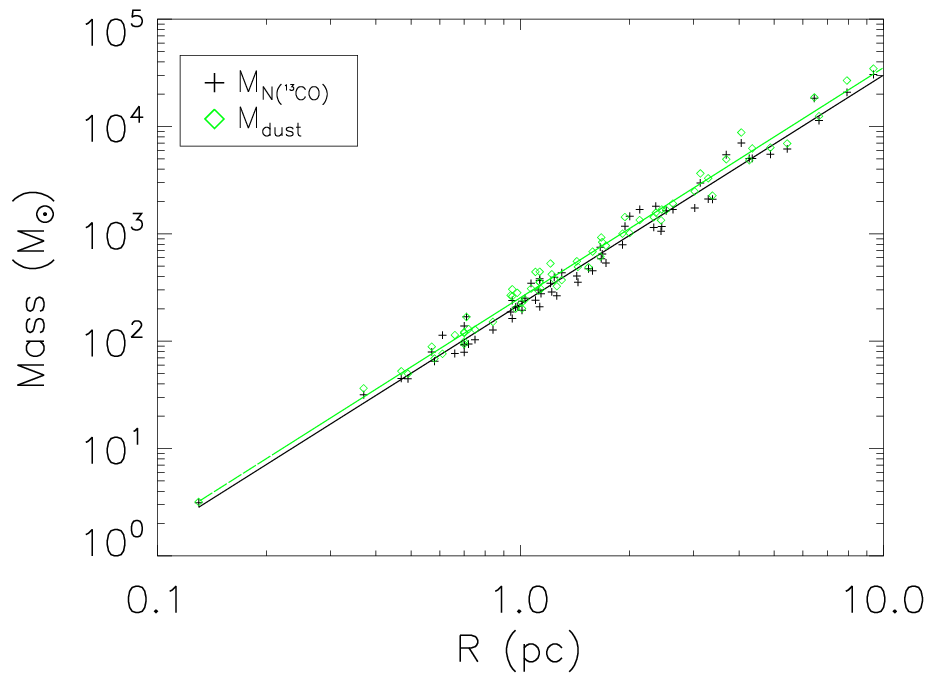}
    \caption{Plot of the the mass of the MCs $\it vs$ their radius. The mass derived from the $^{13}$CO column density is indicated with black pluses, and the mass derived from \herschel\, -based H$_2$ column density is indicated with green diamonds. The linear fit in the log-log axes for the two data-sets is shown by the straight lines.}
   \label{fig:massradius}
\end{figure}

In Fig. \ref{fig:massradius} we show the mass-radius relation for the MCs of the FQS catalogue with masses derived from both the gas (namely FQS data) and dust (\herschel\, data). The masses and radii are well correlated, with a power law of the type $M \propto R^D$ with the exponent $D$ = 2.13$\pm$0.04 for the masses derived from CO and $D$ = 2.14$\pm$0.03 for the masses derived from dust. The two exponents (derived with the \texttt{robust\_linefit} IDL procedure) are consistent within the statistical error and similar to that previously derived in the catalogue of MCs of GRS $D$ = 2.36$\pm$0.04 \citep{roman-duval2010}. The strict correlation between mass and radius with a power-law exponent close to 2 leads to a range in mass surface density where the minimum and maximum values differ by only a factor of $\sim$ 2 with respect to the modal value of 80 \msun\, pc$^{-2}$. From this correlation, we can find the minimum detectable mass that corresponds to the minimum detectable area at the various distances, that is, 10 \msun\, at 1 kpc and 880 \msun\, at 8 kpc. The minimum area and mass at the largest distance ($\sim$ 8 kpc) roughly correspond to the values below which the respective distributions invert the slope (see Fig. \ref{fig:histograms}), which indicates that our sample is incomplete below these values. 

We note that about 60\% of our MCs have a virial parameter smaller than two, which is the critical value below which a structure is gravitationally bound. This indicates that in the majority of the identified clouds, gravity plays a major role in their confinement. 

\subsection{Comparison between the FQS molecular cloud catalogues from \comain\ and \coisot}

In this subsection, we compare the properties of the MCs catalogue derived from the \coisot\, data with those  derived from the \comain\, line in the same region \citep{benedettini2020}. Because the two lines have a slightly different critical density and optical depth, they trace gas with different physical properties and therefore different parts of the same structure. It is therefore interesting to compare the two FQS catalogues to understand how the measured physical parameters of the MCs change in the two catalogues and to determine whether they can lead to different conclusions about the status of the cloud itself, for example, whether they are gravitationally bound structures.

\begin{figure}
   \centering
    \includegraphics[width=8.cm]{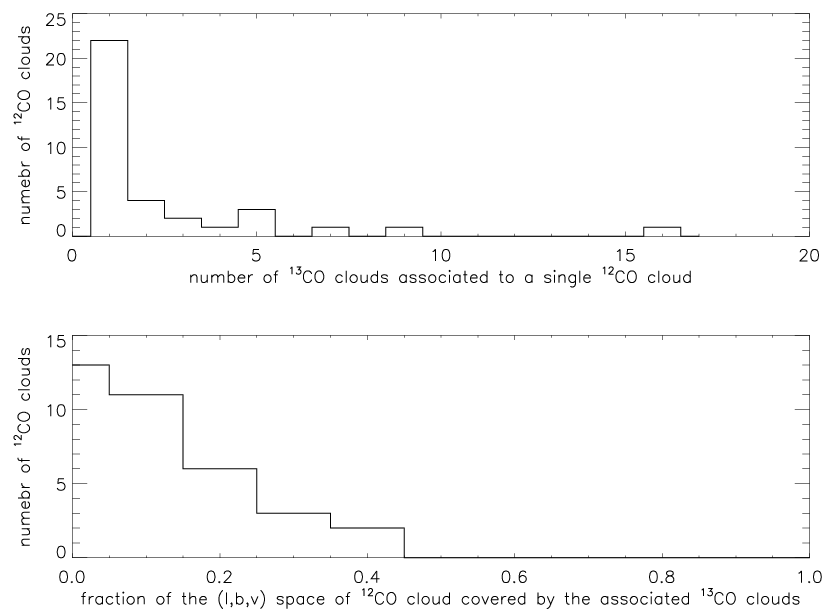}
    \caption{{\it Top:} Histogram of the number of $^{13}$CO MCs associated with a single $^{12}$CO MC. {\it  Bottom:} Histogram of the fraction of the ({\it l,b,$\varv$}) space of $^{12}$CO MC covered by the associated $^{13}$CO MCs.}
   \label{fig:comp1213}
\end{figure}

In our FQS data, the area in which \coisot\, emission is detected is smaller than that in which \comain\, emission is detected because the $^{13}$CO chemical abundance is lower (see Figs. from 6 to 9 in \citealt{benedettini2020}). On the other hand, the lower opacity of \coisot\, allows us to probe a deeper column of gas inside the cloud, with a higher particle density than for the \comain\, line. As a consequence, the structures traced by the \coisot\, line are the brightest and densest parts of the \comain\, structures, while low-brightness regions in \comain\, are mostly undetected in \coisot. Moreover, extended \comain\, MCs with an internal substructure may become decomposed in separate clouds in the \secondcat. We find that 35 out of the 267 MCs of the \firstcat\, (i.e. 13\%) have one or more counterparts in the \secondcat. For these MCs, we show in Fig. \ref{fig:comp1213} the distribution of how many $^{13}$CO MCs are associated with any single $^{12}$CO MC. In general, there is a one-to-one association, but 13 $^{12}$CO MCs are composed of two or more $^{13}$CO MCs. This last group is composed of $^{12}$CO MCs that show substructures, being formed by more than one leaf that are clustered together by the SCIMES algorithm. It is worth noting that in our FQS data, for each $^{12}$CO MC with at least one associated $^{13}$CO MC, the percentage of the ({\it l,b,$\varv$}) cube that has a \comain\, signal above  2$\times$ rms(\comain) and also contains  \coisot\, signal above 3$\times$ rms(\coisot) is low, between 1\% and 47\% (bottom panel of Fig. \ref{fig:comp1213}). This confirms that only the brightest \comain\, emission is also detected in \coisot.

Comparing the median and modal values of the main physical parameters in the two FQS catalogues, we find that in general, the \coisot\, clouds (Table \ref{tab:statistic}) are smaller and less massive than those identified in the \comain\, catalogue (Table 1 in \citet{benedettini2020}), while the surface density of the structures traced by \coisot\, is slightly higher than that of the \comain, as expected. On the other hand, the median and modal values of the aspect ratio ($\sigma_{\rm maj}$ / $\sigma_{\rm min}$), equivalent spherical radius, velocity dispersion and virial parameter are similar in the two catalogues, as well as their parameter range and the shape of the distributions. In particular, it is interesting to compare the values of the velocity dispersion derived from the FWHM of \comain\, and \coisot\, since the first line has a higher optical depth than the second and it is well known that velocity dispersion of the gas derived from the FWHM of an optically thick line tends to be overestimated by a factor that increases with the line opacity \citep{hacar2016}. 

\begin{figure}
   \centering
    \includegraphics[width=9.cm]{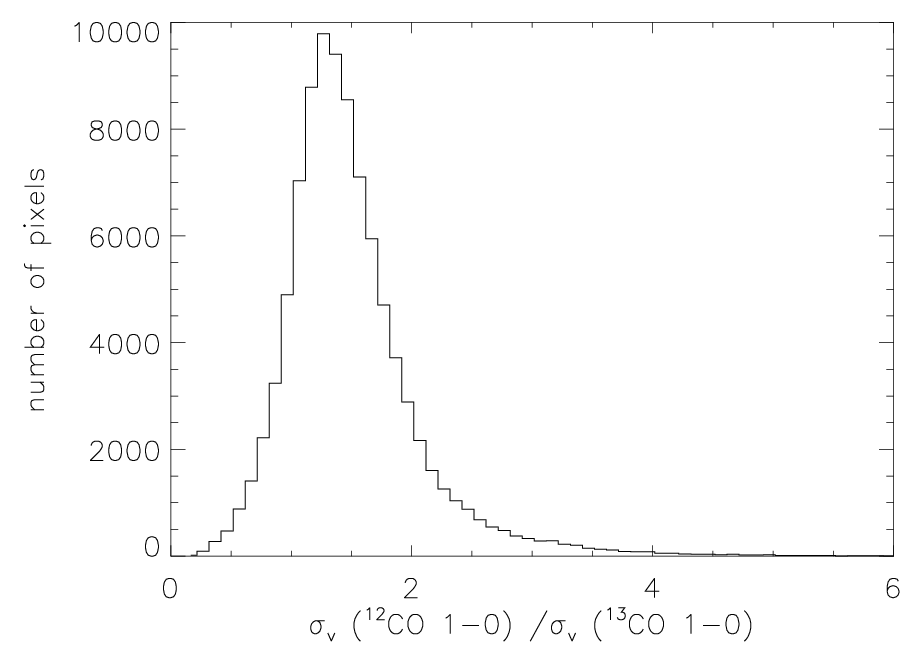}
    \caption{Histogram of the ratio of the velocity dispersion derived from the FWHM of the \comain\, and \coisot\, lines.}
   \label{fig:vel_disp_ratio}
\end{figure}

In Fig. \ref{fig:vel_disp_ratio} we show the ratio of the velocity dispersion derived from the FWHM of the \comain\, and \coisot\, lines in each pixel. As expected, for most of the pixels, the velocity dispersion derived from the line with higher opacity, \comain, is larger. However, the difference is small, with a median ratio of 1.4. This indicates that the velocity dispersion of the MCs of the \firstcat\, of \citet{benedettini2020} is a stringent upper limit of the real velocity dispersion of the gas.
In \citet{benedettini2020} we discussed the relation between the velocity dispersion and the radius of the MCs, known as the first Larson relation \citep{larson1981}, which is expected to be of the form $\sigma_{v} \propto R^\beta$ with $\beta$ = 0.5, if the origin of the gas motions is purely supersonic turbulence, the so-called Burgers turbulence \citep{mckee2007}. For the MCs of the \firstcat,\, we found two regimes: for $R\geq$ 2 pc, the velocity dispersion increases with radius with an exponent $\beta$ = 0.59, similar to what is expected from pure supersonic turbulence, while for $R<$ 2 pc, the relation is almost flat, with $\beta$ = 0.08, similar to what is found in massive clumps \citep{traficante2018a,traficante2018b}, suggesting that gravity already plays an important role at these scales, as also suggested by the values of the virial parameter we estimated. This behaviour was also observed in the MC catalogue derived for the Milky Wave plane in the Galactic longitude range 104\fdg75 $< l <$ 119\fdg75 \citep{ma2021}. 

\begin{figure}
   \centering
    \includegraphics[width=9.cm]{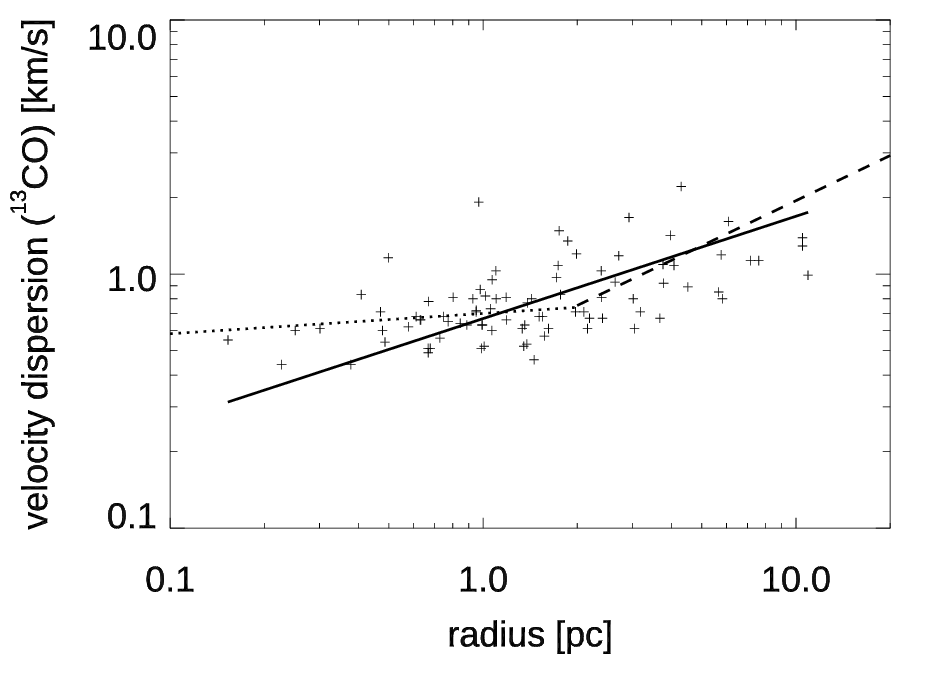}
    \caption{Radius {\it vs} velocity dispersion plot of MCs of the $^{13}$CO catalogue. The solid line represents the best fit for the relation $\sigma_{v} = a R^\beta$ in the full radius range. The best-fit coefficients are $a$ = 0.67 and $\beta$ = 0.4. The dotted and dashed lines correspond to the relations derived in \citet{benedettini2020} for the \firstcat\, rescaled to this set of data: $a$ = 0.7 and $\beta$ = 0.08 for $R<$ 2 (dotted line), and $a$ = 0.5 and $\beta =$ 0.59 for $R>$ 2 (dashed line).}
   \label{fig:vel_disp-radius}
\end{figure}

We repeated this analysis for the MCs detected in this paper, which represent the densest part of some of the largest clouds of \citet{benedettini2020}. The result is shown in Fig. \ref{fig:vel_disp-radius}. The limited number of MCs and the restricted radius range prevent the detection of a clear trend. The observed distribution seems to be compatible with the behaviour found for the \firstcat\, clouds (shown as dotted and dashed lines in Fig. \ref{fig:vel_disp-radius}) and also with an only supersonic turbulent regime, indicated by the solid line that is the best fit of the relation $\sigma_{v} = a R^\beta$ in the full radius range, whose coefficients are: $a$ = 0.67 and $\beta$ = 0.4. The Pearson coefficient is also similar in both cases, 0.4 in the case of two power laws and 0.6 in the case of a single power law. This show a low degree of correlation between the two parameters for the clouds of the \secondcat. 

\begin{figure}
   \centering
    \includegraphics[width=9.cm]{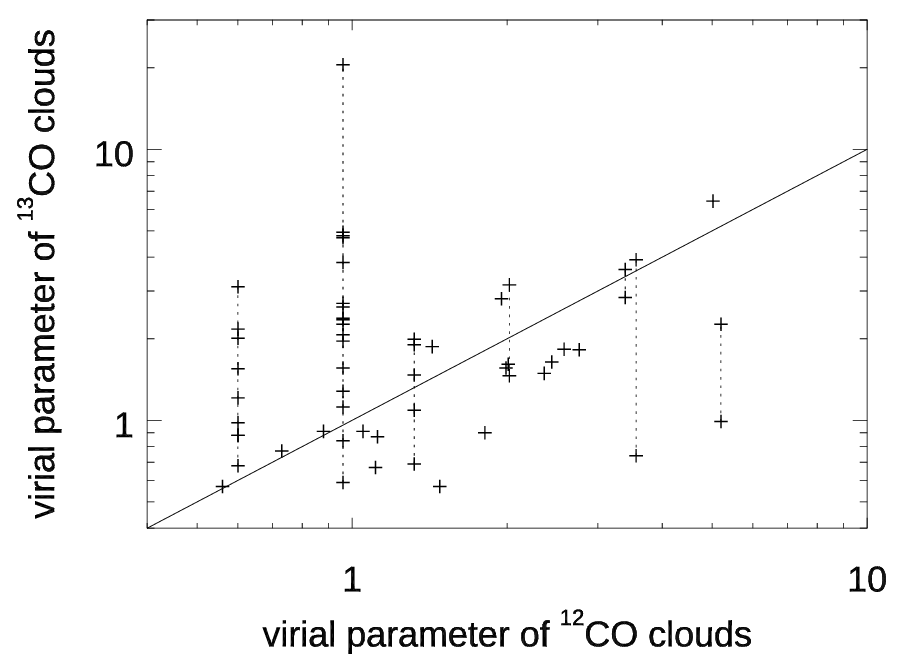}
    \caption{Plot of the virial parameter of the \comain\, MCs that have at least one \coisot\, associated MC, with the virial parameter of the associated \coisot\, MC(s). Only the sub-sample of MCs with a valid determination of the virial parameter (flag1=0 in Table \ref{tab:MC}) are plotted. Dotted vertical lines connect the virial parameters of the $^{13}$CO MCs associated with the same single $^{12}$CO MC. The solid straight line indicates the equivalence of the two parameters.}
   \label{fig:virial1213}
\end{figure}

Another interesting comparison regards the virial parameter derived from the two catalogues, which is used to evaluate the role of gravity in the confinement of the clouds. In Fig. \ref{fig:virial1213} we show the plot of the virial parameter of the \comain\, MCs that have at least one \coisot\, associated MC, against the virial parameter of all the associated \coisot\, MCs. The virial parameters of the \coisot\, clouds associated with the same single \comain\, cloud are connected by a dotted vertical line in Fig. \ref{fig:virial1213}. When we exclude the \comain\, clouds with many \coisot\, sub-structures, the two virial parameters are well correlated. The virial parameter derived from \coisot\, is similar to or slightly lower than that derived from \comain. This confirms that the fraction of the cloud traced by \coisot\, is the densest and hence more gravitationally bound part. We calculated the ratio of the virial parameters derived from the \comain\, with respect to that derived from \coisot for the $^{12}$CO MCs associated with only one $^{13}$CO cloud. We found  a mean value $< \alpha_{\rm vir} [^{12}\rm CO (1-0)]/\alpha_{\rm vir} [^{13}\rm CO (1-0)] >$ = 1.3, that is, a value similar to the mean ratio of the velocity dispersion discussed above. This modest difference likely arises because in the surveyed portion of the Galactic plane, the H$_2$ column density is not particularly high ($\leq$ 5$\times 10^{22}$ \cmdue), leading to a CO emission that does not have a very high optical depth. Moreover, the effects of the different line opacity between the two tracers are small. Our conclusions cannot be generalised  to the entire Milky Way, however, because zones of the Galactic plane with higher H$_2$ column density and particle density may have more different velocity dispersions and virial parameters. 

\section{Surface density versus galactocentric radius}

\begin{figure}
   \centering
    \includegraphics[width=8.7cm]{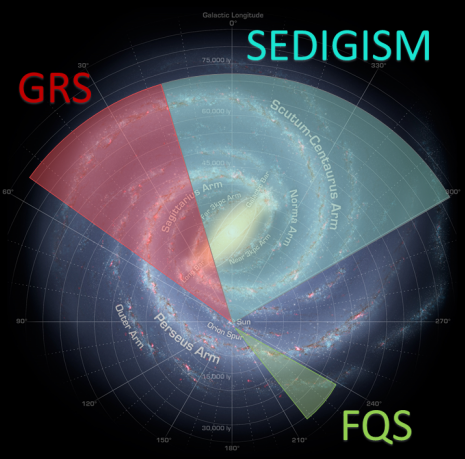}
    \caption{Artistic representation of the Milky Way. The Galactic longitude range covered by  three spectroscopic surveys is indicated with coloured slices: red for GRS \citep{jackson2006}, cyan for SEDIGISM \citep{schuller2017}, and green for FQS \citep{benedettini2020}.}
   \label{fig:mw}
\end{figure}

\begin{figure*}
   \centering
    \includegraphics[width=0.9\textwidth]{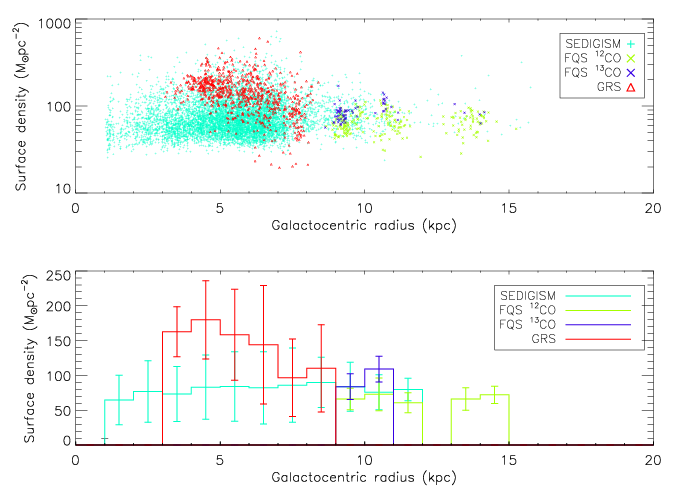}
    \caption{{\it Top:} Plot of the mass surface density {\it vs} galactocentric radius for the MCs of the SEDIGISM catalogue (cyan plus, \citealt{duarte-cabral2021}), FQS \firstcat\, (green crosses, \citealt{benedettini2020}), FQS \secondcat\, (blue crosses, this paper), and GRS (red triangles, \citealt{roman-duval2010}. {\it Bottom:} Histograms of the mean value of the data shown in the top panel in 1 kpc bins, for bins with more than ten sources. The error bars are the statistical standard deviations in each bin. }
   \label{fig:surfdens_rgal}
\end{figure*}

One of the  essential but not sufficient conditions for activating star formation inside an MC is that enough mass must be gathered in some part of the cloud. An easy observable of this property is the mass surface density. The variation in mass surface density of the MCs with galactocentric radius has been studied by several authors \citep{miville2017, roman-duval2010}, who have found a decline in molecular cloud mass surface density at $R_{\rm gal}\gtrsim$ 6 kpc. Interestingly, a decrease in mass surface density from the inner to the outer Galaxy was also observed  for protostars and pre-stellar cores/clumps \citep{elia2021}, indicating a lower star formation activity in the outer  portion of the Milky Way. We used our FQS MC catalogues together with catalogues produced from homogeneous data of new-generation spectral surveys to study the variation in surface density of the MCs across the Galaxy in a large galactocentric radius range. In particular, we used the \citet{roman-duval2010} catalogue, extracted from the GRS data (line \coisot, Galactic longitude range 18\degr $< l <$ 55\fdg7, angular resolution 46\arcsec, velocity resolution 0.21 \kms, $R_{\rm gal}$ = 3 -- 8 kpc, and extractor algorithm CLUMPFIND) and the \citet{duarte-cabral2021} catalogue based on the SEDIGISM data (line $^{13}$CO (2--1), Galactic longitude range -60\degr $< l <$ 18\degr, angular resolution 30\arcsec, velocity resolution 0.25 \kms, $R_{\rm gal}$ = 1 -- 15 kpc, and extractor algorithm SCIMES). In Fig. \ref{fig:mw} we show the Galactic longitude range covered by the three considered surveys: FQS, GRS, and SEDIGISM, which corresponds to $\sim$ 40\% of the total range.

The mean value of the mass surface density of the MCs of the three catalogues is 87$\pm$55 \msun\, pc$^{-2}$, which can be considered  as the reference value for the Milky Way. We note that the mean value derived from FQS data perfectly agrees with the total mean value.
In Fig. \ref{fig:surfdens_rgal} we show the relation of the mass surface density of MCs with $R_{\rm gal}$, together with the histograms of the surface density mean value in 1 kpc bins. In order to have a minimum statistical value, we considered only bins with more than ten sources. For the MCs of the FQS \firstcat\, (green symbols) and \secondcat\, (blue symbols), we find that the mean value of the mass surface density is constant in the range $\sim$ 8 -- 16 kpc, with a slightly higher value for the MCs of the \secondcat\, compared to those of the \firstcat. Our FQS data agree with the flat trend of the mean mass surface density of the SEDIGISM catalogue (cyan symbols), covering both the inner and outer Galaxy, and  at $R_{\rm gal}$ = 9 kpc connect well with the trend of the GRS data, which start to decrease at $R_{\rm gal} \gtrsim$ 6 kpc. Conversely, in the inner Galaxy, the GRS  mean values are higher than the SEDIGISM mean values, and in the range 3 $\lesssim R_{\rm gal} \lesssim$ 6 kpc, they are not consistent at the  $1\sigma$ level. Interestingly, we find a variability range of the MC mass surface density of about one order of magnitude in the whole range of the investigated Galactocentric radii. The highest values of the SEDIGISM clouds are compatible with the values of the GRS clouds (see Fig. \ref{fig:surfdens_rgal}). This is an indication that this parameter, which is strictly related to the capability of a cloud to form stars, is determined more by local conditions than by the distance of the cloud from the centre of the Galaxy. 

It is worth noting that the measured mass surface density of the MCs can be affected by the crowding level of clouds. Indeed, in highly crowded regions, such as the central part of the Galaxy and the tips of the bar, it is more difficult to properly separate structures along the same line of sight. This leads to a possible overestimate of the cloud mass and consequently of the mass surface density. This effect might explain the prevalence in GRS data of a mass surface density higher than $\sim$ 100 \msun\, pc$^{-2}$ (see Fig. \ref{fig:mw}). The GRS spans the Galactic longitude range 18\degr $< l <$ 55\fdg7, which is largely covered by the tip of the bar, while SEDIGISM covers a much wider range -60\degr $< l <$ 18\degr, which includes part of the bar with the other tip, but also lines of sight that do not cross the central region of the Galaxy and therefore are less crowded. FQS includes only the outer Galaxy with a low level of crowding. As demonstrated by \citet{veneziani2017}, the tips of the bar show an enhanced star formation rate with respect to background and foreground regions because of the large amount of dust and molecular material they contain. This can contribute to the higher mean mass survey density of MCs measured in the GRS catalogue. 

We must consider that the sensitivity level and the method we used to derive the cloud mass is different for the three catalogues. The GRS survey has a sensitivity of about a factor  2 better than FQS and a factor 5 better than SEDIGISM, although it observes the $^{13}$CO (2-1) line. To derive mass, SEDIGISM used the same method as described in Sect. \ref{sect:mass_xco}, with the CO-to-H$_2$ conversion factor $X(^{13}{\rm CO (2-1)})$ = 10$^{21}$ cm$^{-2}$ (K km s$^{-1}$)$^{-1}$, derived from the comparison with the \herschel\, H$_2$ column density maps \citep{schuller2017}. The mass surface density in our FQS catalogue is calculated from the mass derived from the \herschel\, data, as described in Sect. \ref{sect:mass_dust}. This means that both FQS and SEDIGISM used the \herschel\, H$_2$ column density as a mass calibrator. On the other hand, the cloud mass in GRS is derived from the $^{13}$CO column density multiplied with the $^{13}$CO chemical abundance of 1.77$\times$10$^{-6}$, which is the method described in Sect. \ref{sect:cd_co}, but with a different abundance. Finally, while the FQS and SEDIGISM catalogues are produced with the same cloud identification algorithm, SCIMES, the GRS catalogue used a different algorithm,  CLUMPFIND. A comparison of the performance of the two algorithms in detecting MCs in the same region was made by \citet{colombo2015}, who found that the two algorithms perform differently and that CLUMPFIND tends to overdivide the molecular emission into smaller structures with respect to SCIMES. In conclusion, the FQS and SEDIGISM catalogues are more homogeneous, while a direct comparison with the GRS catalogue requires more caution.

\section{Conclusions}

We presented a catalogue of MCs extracted from the $^{13}$CO (1--0) data cubes at a spectral resolution of 1 \kms\, of the FQS survey, covering the Galactic plane in the range 220\degr$<l<$240\degr\, and -2\fdg5$<b<$0\degr. The catalogue contains 87 MCs, for which the main physical parameters, such as area, mass, distance, velocity dispersion and virial parameter, were derived.
Making use of available ancillary data, that is, the FQS \comain\, emission and the \herschel\, $N({\rm H_2})$ column density map, we applied three different methods to derive the MC mass and compared the results. In general, we find a good agreement between the three different mass estimates. In particular, the masses derived from CO data with $N(\rm ^{13}CO)$ and $X(\rm^{12}CO)$ are almost similar, with a narrow distribution of their ratio centred at the median value of $<M_{N(^{13}\rm CO)} / M_{X(^{12}{\rm CO})}>$ = 0.99, indicating that the less extended \coisot\, emission loses only the more tenuous outskirts of the clouds, which contribute little to the total cloud mass. On the other hand, for most of the MCs, the mass derived from the CO gas is lower than that derived from the cold dust, with median values of $<M_{N(^{13}\rm CO)} / M_{\rm dust}>$ = 0.86 and $<M_{X(^{12}{\rm CO})} / M_{\rm dust}>$ = 0.86. In limited areas, we find that $M_{N(^{13}\rm CO)} / M_{\rm dust} \geq$ 2, likely due to a modification of the CO chemical abundance and/or dust grain emissivity related to the high temperature and ionisation level of these regions. 

We compared the physical parameters of the MCs of the FQS catalogue extracted from \coisot\, with those derived from \comain\, presented in \citet{benedettini2020} and found that the structures traced by the \coisot\, emission are associated with the brightest and densest regions of the \comain\, structures, while the structures faint in \comain\, remain undetected in \coisot. In particular, we found that 35 out of the 267 MCs of the \firstcat\, have at least one counterpart in the \secondcat. For these clouds, not more than 50\% of the ({\it l,b,$\varv$}) cube with \comain\, emission is also covered by \coisot\, emission. As a consequence, the \coisot\, MCs are smaller and less massive than the \comain\, MCs, while the mass surface density of the structures traced by \coisot\, is slightly higher than the mass traced in \comain. On the other hand, the median and modal values of the aspect ratio, equivalent spherical radius, velocity dispersion and virial parameter are similar in the two catalogues, as well as their parameter range and the shape of their distributions. This is likely due to the fact that in the surveyed portion of the Galactic Plane the H$_2$ column density is not particularly high ($\leq$ 5$\times 10^{22}$ \cmdue) leading to a CO emission that has not a very high optical depth. This reduces the effects of the different line opacity between the two tracers on some of the estimated physical parameters.

By complementing our FQS MC catalogues with those of the GRS and SEDIGISM surveys, we derived a mean value of the cloud mass surface density of 87$\pm$55 \msun\, pc$^{-2}$ in the Milky Way. Even though this value spans about an order of magnitude over the single clouds, its mean value in 1 kpc bins is almost constant for a galactocentric radius between 3 kpc and 16 kpc, indicating that this physical property is affected more by local conditions than by the distance of the cloud from the centre of the Galaxy. 
The comparison of the results from different MC catalogues, however, must be performed with caution because the measured physical properties depend not only on the molecular tracer, but also on the method with which the parameters were derived, in particular, the mass, and on the algorithm with which the structures were identified.

\begin{acknowledgements}
We thank D. Colombo for his valuable support in the use of the SCIMES algorithm.
This research is supported by INAF, through the Mainstream Grant 1.05.01.86.09 ``The ultimate exploitation of the Hi-GAL archive and ancillary infrared/mm data''.
\end{acknowledgements}

\begin{appendix}

 \section{Physical properties of the \secondcat}
 In Table \ref{tab:MC} we list all the physical parameters derived for the 87 MCs extracted from the FQS \coisot\, spectral cube. Only a few objects are listed as examples in the table. The complete catalogue is available as online material at the Centre de Donn\'{e}es astronomiques de Strasbourg (CDS). The description of the columns is reported in the footnote of the table.

\begin{table*}
\caption{Properties of the molecular clouds identified in the \coisot\, data cubes. Only a few objects are listed as examples here. The complete catalogue is available in digital form at the CDS.}             
\label{tab:MC}      
\centering

\begin{tabular}{r c c c r r r r r r }     
\hline\hline    
index & Name & $l$ & $b$ & $\sigma_{\rm maj}$ & $\sigma_{\rm min}$ & PA & $\varv_{\rm lsr}$ & $\sigma_{v}$ & $I({\rm ^{13}CO})$  \\
            & & \degr& \degr& \arcsec &\arcsec & \degr & \kms & \kms& K \kms  \\
\hline                    
1  & FQS-MC219.860-2.201  & 219.8598  &  -2.2009  &  43.51  &  31.49 & 102.12 &  11.75  &  0.61  &     3.68 \\
2  & FQS-MC220.081-2.160  & 220.0811  &  -2.1603  &  176.67 &  92.63 &  53.25 &  12.07  &  0.60  &     2.93 \\
5  & FQS-MC220.321-1.751  & 220.3213  &  -1.7506  & 117.33  &  57.61 &  57.01 &  12.80 &   0.56 &      3.18 \\
8  & FQS-MC220.716-1.783 &  220.7156  &  -1.7830  & 453.66  & 188.49 & 122.42 &  11.90 &   1.03 &      5.33 \\
12 & FQS-MC221.283-1.758 &  221.2829  &  -1.7578  &  59.98  &  29.75 &  56.32 &   5.45 &   0.55  &     3.20 \\
16 &  FQS-MC222.755-1.644 &  222.7553 &   -1.6441 & 436.88  & 221.78 & 172.00 &  17.06 &   0.67 &      3.01 \\
\hline           
\end{tabular}
\\
\vspace{0.5cm}
\begin{tabular}{r r r r r r r r r r r}
\hline\hline       
$d$ & $R$ & $A$ & $M_{N(^{13}\rm CO)}$ & $M_{X(^{12}{\rm CO})}$ & $M_{\rm dust}$ & $\Sigma$ & $\alpha_{\rm vir}$ &  n$_{\rm l}$ & flag1 & flag2 \\
kpc & pc & pc$^2$ & \msun & \msun & \msun & \msun\, pc$^{-2}$ \\
\hline       
 0.877 &    0.301 &   0.32 &     24.37 &     13.12 &      0.00 &   0.00 &     0.00  &    1 &     1 &     2  \\
 0.900 &    1.067 &   2.35 &    131.53 &     97.91 &      7.12 &   3.03 &    63.15 &    1 &     1 &     0  \\
 0.956 &    0.728 &   1.56 &     92.31 &     71.79 &    119.05 &  76.48 &     2.23 &     1 &     0 &     0  \\
 0.882 &    2.387 &  17.43 &   1813.60 &   1714.00 &   1574.20 &  90.25 &     1.88 &     3 &     0  &    0 \\
 0.391 &    0.153 &   0.06 &      3.12 &      4.10 &      3.18 &  56.56 &    17.09 &     1 &     0  &    2 \\
 1.275 &    3.675 &  18.77 &   1059.70 &   1199.00 &   1337.50 &  71.10 &     1.46 &     2 &     0  &    0 \\
\hline                    
\end{tabular}

\tablefoot{Columns are as follow. Column 1; progressive index. Column 2; name defined as ``FQS-MC'' followed by the Galactic coordinates of the cloud centroid. Columns 3 and 4: Galactic longitude and latitude of the cloud centroid. Column 5: intensity-weighted semi-major axis. Column 6: intensity-weighted semi-minor axis. Column 7: position angle w.r.t. the cube x-axis. Column 8: mean velocity. Column 9: velocity dispersion. Column 10: \coisot\, integrated intensity across the area of the cloud. Column 11: kinematic distance from the Sun. Column 12: equivalent spherical radius. Column 13: area. Column 14: total mass derived from $^{13}$CO column density. Column 15: total mass derived from \comain\, line intensity multiplied for the $X(^{12}{\rm CO})$ factor. Column 16: total mass derived from \herschel\, H$_2$ column density. Column 17: average surface density. Column 18: virial parameter. Column 19: number of dendrogram leaves. Column 20: flag1; 0 indicates that the cloud is fully mapped in the \herschel\, H$_2$ column density map, 1 indicates that the area of the cloud as derived from the \coisot\, data is only partially covered or not covered at all by the \herschel\, H$_2$ column density map; for those clouds $M_{\rm dust}$ and $\Sigma$ are lower limits and $\alpha_{\rm vir}$ is an upper limit. Column 21: flag2; 0 means that the cloud is fully mapped in \coisot, 2 means that the area of the cloud as derived from \coisot\, reaches the edge of the map; these clouds could extend outside the mapped area, therefore the measured parameters show an uncertainty that depends on how much CO emission was missed.}

\end{table*}

\end{appendix}


\begin{thebibliography}{}
\bibitem[\protect\citeauthoryear{Barnes et al.}{2015}]{barnes2015}
Barnes, P. J., Muller, E., Indermuehle, B., et al. 2015, ApJ, 812, 6
\bibitem[\protect\citeauthoryear{Paper I}{}]{benedettini2020}
Benedettini, M., Molinari, S., Baldeschi, A., et al. 2020, A\&A, 633, A147, Paper I
\bibitem[\protect\citeauthoryear{Bolatto et al.}{2013}]{bolatto2013}
Bolatto, A.D., Wolfire, M., \& Leroy, A.K. 2013, ARA\&A, 51, 207
\bibitem[\protect\citeauthoryear{Burton et al.}{2013}]{burton2013}
Burton, M. G., Braiding, C., Glueck, C., et al. 2013, PASA, 30, 44
\bibitem[\protect\citeauthoryear{Choi et al.}{2014}]{choi2014}
Choi, Y. K., Hachisura, K., Reid, M. J., et al. 2014, ApJ, 790, 99
\bibitem[\protect\citeauthoryear{Colombo et al.}{2015}]{colombo2015}
Colombo, D., Rosolowsky, E., Ginsburg, A., Duarte-Cabral, A., \& Hughes, A. 2015, MNRAS, 454, 2067
\bibitem[\protect\citeauthoryear{Dame et al.}{2001}]{dame2001}
Dame, T. M., Hartmann, D. \& Thaddeus, P. 2001, ApJ, 547, 792
\bibitem[\protect\citeauthoryear{Dempsey et al.}{2013}]{dempsey2013}
Dempsey, J. T., Thomas H. S., Currie M. J., 2013, ApJS, 209, 8
\bibitem[\protect\citeauthoryear{D\'{e}sert et al.}{2008}]{desert2008}
D\'{e}sert, F., Mac\'{i}as-P\'{e}rez, J. F., Mayet, F., et al. 2008, A\&A, 481, 411
\bibitem[\protect\citeauthoryear{Duarte-Cabral et al.}{2021}]{duarte-cabral2021}
Duarte-Cabral, A., Colombo, D., Urquhart, J. S., et al. 2021, MNRAS, 500, 3027
\bibitem[\protect\citeauthoryear{Dupac et al.}{2003}]{dupac2003}
Dupac, X., Bernard, J., Boudet, N., et al. 2003, A\&A, 404, L11
\bibitem[\protect\citeauthoryear{Elia et al.}{2013}]{elia2013}
Elia, D., Molinari, S., Fukui, Y., et al. 2013, ApJ, 772, 45
\bibitem[\protect\citeauthoryear{Elia et al.}{2021}]{elia2021}
Elia, D., Merello, M., Molinari, S., et al. 2021, MNRAS, 504, 2742
\bibitem[\protect\citeauthoryear{Fernandes et al.}{2019}]{fernandes2019}
Fernandes, B., Montmerle, T., Santos-Silva, T. \& Gregorio-Hetem, J. 2019, A\&A 628, A44
\bibitem[\protect\citeauthoryear{Frerking et al.}{1982}]{frerking1982}
Frerking, M., Langer, W. D., \& Wilson, R. W., 1982, ApJ, 262, 590
\bibitem[\protect\citeauthoryear{Giannetti et al.}{2017}]{giannetti2017}
Giannetti, A., Leurini, S., K\"{o}nig, C., et al. 2017, A\&A, 606, L12
\bibitem[\protect\citeauthoryear{Hacar et al.}{2016}]{hacar2016}
Hacar, A., Alves, J., Burkert, A., \& Goldsmith, P. 2016, A\&A, 591, A104
\bibitem[\protect\citeauthoryear{Hildebrand}{1983}]{hildebrand1983}
Hildebrand R. H., 1983, QJRAS, 24, 267
\bibitem[\protect\citeauthoryear{Hou \& Han}{2014}]{hou2014}
Hou, L. G. \& Han, J. L. 2014, A\&A, 569, A125
\bibitem[\protect\citeauthoryear{Jackson et al.}{2006}]{jackson2006}
Jackson, J.M., Rathborne, J.M., Shah R.Y., et al. 2006, ApJS, 163, 145
\bibitem[\protect\citeauthoryear{Jiang \& Li}{2013}]{jiang2013}
Jiang, Z. \& Li, J. 2013 in Protostars and Planets VI, Conference Poster n. 1B003
\bibitem[\protect\citeauthoryear{Juvela et al.}{2015}]{juvela2015}
Juvela, M., Demyk, K., Doi, Y., et al. 2015, A\&A 584, A94
\bibitem[\protect\citeauthoryear{Larson}{1991}]{larson1981}
Larson, R.B. 1981, MNRAS, 194, 809
\bibitem[\protect\citeauthoryear{Lin et al.}{2021}]{lin2021}
Lin, Z., Sun, Y., Xu, Y., et al. 2021, ApJS, 252, 20
\bibitem[\protect\citeauthoryear{Ma et al.}{2021}]{ma2021}
Ma, Y., Wang, H., Li, C., et al. 2021, ApJS, 254, 3
\bibitem[\protect\citeauthoryear{Martin et al.}{2012}]{martin2012}
Martin, P. G., Roy, A., Bontemps, S., et al. 2012, ApJ, 751, 28
\bibitem[\protect\citeauthoryear{McKee \& Zweibel}{1992}]{mckee1992}
McKee, C.F. \& Zweibel, E.G. 1992, ApJ, 399, 551
\bibitem[\protect\citeauthoryear{McKee \& Ostricker}{2007}]{mckee2007}
McKee, C.F. \& Ostricker, E.C. 2007, ARAA, 45, 565
\bibitem[\protect\citeauthoryear{Miville-Desch\^{e}nes et al. }{2017}]{miville2017}
Miville-Desch\^{e}nes, M.A., Murray, N., \& Lee, E.J. 2017, ApJ, 834, 57
\bibitem[\protect\citeauthoryear{Mottram \& Brunt}{2010}]{mottram2010}
Mottram, J. C. \& Brunt, C. M. 2010, in SP Conf. Ser. 438, The Dynamic Interstellar Medium: A Celebration of the Canadian Galactic Plane Survey, ed. R. Kothes, T. L. Landecker, and A. G. Willis, p. 98
\bibitem[\protect\citeauthoryear{Olmi et al.}{2016}]{olmi2016}
Olmi, L., Cunningham, M., Elia, D., \& Jones, P. 2016, A\&A, 594, A58
\bibitem[\protect\citeauthoryear{Pezzuto et al.}{2021}]{pezzuto2021}
Pezzuto, S., Benedettini, M., Di Francesco, J., et al. 2021, A\&A, 645, A55
\bibitem[\protect\citeauthoryear{Reid et al.}{2014}]{reid2014}
Reid, M., Menten, K.M., Brunthaler, A., et al. 2014, ApJ, 783, 130
\bibitem[\protect\citeauthoryear{Rigby et al.}{2016}]{rigby2016}
Rigby, A.J., Moore, T.J.T., Plume, R., et aal. 2016, MNRAS, 456, 2885
\bibitem[\protect\citeauthoryear{Roman-Duval et al.}{2010}]{roman-duval2010}
Roman-Duval, J., Jackson, J.M., Heyer, M., Rathborne, J., \& Simon, R. 2010, ApJ, 723, 492 
\bibitem[\protect\citeauthoryear{Rosolowsky \& Leroy}{2006}]{rosolowsky2006}
Rosolowsky, E. W., \& Leroy, A. 2006, PASP, 118, 590
\bibitem[\protect\citeauthoryear{Roy et al.}{2013}]{roy2013}
Roy, A., Martin, P. G., Polychroni, D., et al. 2013, ApJ, 763, 55
\bibitem[\protect\citeauthoryear{Schisano et al.}{2020}]{schisano2020}
Schisano, E., Molinari, S., Elia, D., et al. 2020, MNRAS, 492, 5420
\bibitem[\protect\citeauthoryear{Schuller et al.}{2017}]{schuller2017}
Schuller, F., Csengeri, T., Urquhart, J. S., et al. 2017, A\&A, 601, A124
\bibitem[\protect\citeauthoryear{Sharpless}{1959}]{sharpless1959}
Sharpless, S. 1959, ApJS, 4, 257
\bibitem[\protect\citeauthoryear{Traficante et al.}{20018a}]{traficante2018a}
Traficante, A., Fuller, G.A., Smith, R.J., et al. 2018a, MNRAS, 473, 4975
\bibitem[\protect\citeauthoryear{Traficante et al.}{2018b}]{traficante2018b}
Traficante, A., Duarte-Cabral, A., Elia, D., et al. 2018b, MNRAS, 477, 2220
\bibitem[\protect\citeauthoryear{Umemoto et al.}{2017}]{umemoto2017}
Umemoto, T., Minamidani, T., Kuno, N., et al. 2017, PASJ, 69, 78
\bibitem[\protect\citeauthoryear{Veneziani et al.}{2017}]{veneziani2017}
Veneziani, M., Schisano, E., Elia, D., et al. 2017, A\&A, 599, A7
\bibitem[\protect\citeauthoryear{Williams et al.}{1994}]{williams1994}
Williams, J. P., de Geus, E. J., \& Blitz, L. 1994, ApJ, 428, 693
\bibitem[\protect\citeauthoryear{Xu et al.}{2013}]{xu2013}
Xu, Y., Li, J. J., Reid, M. J., et al. 2013, ApJ, 769, 15
\end{thebibliography}
\end{document}